\documentclass[twocolumn,showpacs,preprintnumbers,amsmath,amssymb]{revtex4}
\usepackage{graphicx}% Include figure files

\usepackage{dcolumn}% Align table columns on decimal point
\usepackage{bm}% bold math
\usepackage{verbatim} % use comment
\usepackage{units}%use nicefrac
\usepackage{tipa}

\newcommand{\be}{\begin{equation}}
\newcommand{\ee}{\end{equation}}
\newcommand{\bea}{\begin{eqnarray}}
\newcommand{\eea}{\end{eqnarray}}

\newcommand{\one}{|1\rangle}
\newcommand{\two}{|2\rangle}

\newcommand{\nupa}{\nu_{\mathrm{PA}}}

\begin{document}
\date{\today}
%\flushbottom \draft
\title{High resolution photoassociation spectroscopy of the $^{6}$Li$_2$ $A(1^1\Sigma_u^+)$ state}
\author{Will Gunton$^1$, Mariusz Semczuk$^1$, Nikesh S. Dattani$^2$, and Kirk W.~Madison$^1$}

\affiliation{$^{1}$Department of Physics and Astronomy, University of British Columbia, Vancouver, Canada\\
$^{2}$ Department of Chemistry, University of Oxford, Oxford, UK\\
}

\begin{abstract}
We present spectroscopic measurements of seven vibrational levels $v=29-35$ of the $A(1^1\Sigma_u^+)$ excited state of Li$_2$ molecules by the photoassociation of a degenerate Fermi gas of  $^6$Li atoms. The absolute uncertainty of our measurements is $\pm 0.00002$~cm$^{-1}$ ($\pm 600$ kHz) and we use these new data to further refine an analytic potential for this state. This work provides high accuracy photo-association resonance locations essential for the eventual high resolution mapping of the $X(1^1\Sigma_g^+)$ state enabling further improvements to the $s$-wave scattering length determination of Li and enabling the eventual creation of ultra-cold ground state $^6$Li$_2$ molecules. 
\end{abstract}


\maketitle

Since the development of laser cooling techniques for atoms, photoassociation spectroscopy (PAS) has been used to make precise measurements of molecular vibrational levels including very weakly bound states that are difficult to access with traditional bound-bound molecular spectroscopy.  The precision of these types of measurements is due in large part to the extremely low ensemble temperatures achievable with laser cooling.  At sufficiently low temperatures the collision energy is so small that the inhomogeneous broadening of the spectral lines can be negligible compared to the natural linewidth.  In addition, cold atomic ensembles can be confined in a very weak optical dipole trap providing extremely long interrogation times and exceedingly small ac Stark shifts from both the confining laser light and the photoassociation light.  Single-color PAS of excited molecular states has allowed for accurate determinations of atomic lifetimes and two-color PAS of ground molecular states has enabled the precise determination of atom-atom scattering lengths and the production of ultra-cold molecules as discussed in several excellent review articles \cite{RevModPhys.71.1, J.Mol.Spectrosc.195, AdvAtMolOptPhys.42, AdvAtMolOptPhys.47, RevModPhys.78.483}.

In this work, we measure the binding energies of seven vibrational levels $v=29-35$ of the $A(1^1\Sigma_u^+)$ excited state of $^{6}$Li$_2$ molecules with an absolute uncertainty of $\pm 0.00002$~cm$^{-1}$ ($\pm 600$ kHz) by photoassociating a quantum degenerate Fermi gas of lithium atoms held in a shallow optical dipole trap.  These measurements further refine the binding energies for these levels extracted from previous studies of various levels in the range $v=0-85$ $^{6,6}$Li$_2$ \cite{Linton1996340,Wang1998,Wang2002,Adohi-Krou2004,LeRoy2009}, and they complement other high resolution PAS measurements of the $v=62-88$ levels \cite{abraham:7773}, while having an absolute uncertainty that is nearly two orders of magnitude smaller.

This measurement is part of a larger project to characterize the excited $A(1^1\Sigma_u^+)$ (this paper) and $c(1^{3}\Sigma_{g}^{+})$ \cite{PhysRevA.87.052505} molecular levels of $^{6}$Li$_2$, as well as the ground spin-singlet $X(1^1\Sigma_g^+)$ and lowest lying spin-triplet $a(1^{3}\Sigma_{u}^{+})$ potentials. In particular, this work is motivated by several factors. First, the levels measured in this work can serve as convenient intermediate states for a high resolution study of the $X(1^1\Sigma_g^+)$ state which is unexplored in the ultra-cold regime, with the exception of experiments involving Feshbach molecules created using the narrow $^6$Li Feshbach resonance near 543~G, resulting from the $v=38$ vibrational level of the $X(1^1\Sigma_g^+)$ state \cite{PhysRevLett.91.080406,PhysRevLett.108.045304}. Second, the determination of the excited state levels is a vital step towards the creation of ground state molecules using a two-photon STIRAP (stimulated Raman adiabatic passage) process \cite{K.-K.Ni10102008}. Finally, the $s$-wave scattering length (a single parameter which describes elastic interactions in the low temperature regime) is most sensitive to the location of the least-bound vibrational levels of the  $X(1^1\Sigma_g^+)$ and $a(1^{3}\Sigma_{u}^{+})$ states. This work provides high accuracy photo-association resonance locations essential for the eventual high resolution mapping (with an uncertainty of less than 100~kHz) of the least bound levels of the $X(1^1\Sigma_g^+)$ state using at atom-molecule dark state, as demonstrated in Rb \cite{PhysRevLett.95.063202} and metastable He \cite{PhysRevLett.96.023203}. 

Our apparatus and experimental method is described in detail in \cite{PhysRevA.87.052505,arXiv:1307.5445}. Briefly, for these photoassociation measurements, we prepare our ensemble by loading a magneto-optical trap (MOT) with $2\times10^7$ $^6$Li atoms and transfering them into a crossed optical dipole trap (CDT). After multiple forced evaporation stages, the ensemble is composed of $4\times10^4$ atoms with equal populations in the $\one \equiv |F=1/2, m_F=1/2\rangle$ and $\two \equiv |F=1/2, m_F=-1/2\rangle$ states, at a temperature (verified by time-of-flight expansion) of $800~$nK.

After this preparation stage, we apply a homogenous magnetic field that cancels any residual field remaining at the location of the atoms. The background field after cancelation is verified to be less then 20~mG via RF spectroscopy between the $F=1/2$ and $F=3/2$ ground hyperfine levels of $^6$Li. The PA light is derived from a single-frequency, tunable Ti:sapphire laser and illuminates the atomic cloud for $2~$s. The PA beam propagates co-linearly with one of the arms of the CDT and is focused to a waist ($1/e^2$ intensity radius) of $50 \, \mu$m. When the photon energy, $h \nupa$, equals the energy difference between the unbound state of a colliding atomic pair and a bound molecular excited state, excited-state molecules form. This process produces atom loss from the trap, which occurs either because the excited state molecule decays into a ground state molecule that we do not detect, or into two free atoms in the unbound continuum with enough energy that they are lost from the CDT. The number of atoms remaining in the CDT is determined by an absorption image of the cloud. The Ti:Sapphire laser is stabilized to a fiber based, self referenced frequency comb \cite{Mills:09,PhysRevA.87.052505} and after stabilization, it has an absolute frequency uncertainty of $\pm 600~$kHz.

%%%%%%%%%%%%%%%%%%%%%%%%%%%%%%%%%
%% OBSERVATIONS / SYSTEMATICS  %%
%%%%%%%%%%%%%%%%%%%%%%%%%%%%%%%%%
%
We observe the PA spectrum for seven vibrational levels ($v'=29-35$) of the $A^{1}\Sigma_{u}^{+}$ state which arise from $s$-wave collisions between atoms in states $\one$ and $\two$, see Table \ref{tab:N1-resonances}. To reduce thermal broadening and the inhomogeneous ac Stark shift produced by the CDT potential, these data were obtained in a low intensity trap ($I_{\mathrm{CDT}} = 9.6~$kW~cm$^{-2}$). At this low trap power, the ensemble temperature is 800~nK ($T/T_{\mathrm{F}} = 0.4$). In order to fully characterize the systematic shifts of the resonance location due to the CDT and PA laser we varied the CDT intensity from 9.6 to 55~kW~cm$^{-2}$ and the PA laser intensity from 65 to 760~kW~cm$^{-2}$ for each of the seven vibrational levels. 

Assuming that the ac Stark shifts produced by the CDT and PA laser are independent and the shift of the resonance position is proportional to the laser intensity, the shift of the resonance locations can be fit to extract the shift rate due to the CDT and PA beams (Table \ref{tab:N1-resonances}). Using the shift rate, the location of the PA resonances under field free conditions can be inferred (Table \ref{tab:binding_energies}). For the $v=30$ vibrational level, we observe a splitting of the resonance into two disctinct loss features separated by $18~$MHz when the CDT intensity is increased from $9.6~$kW~cm$^{-2}$ to 55~kW~cm$^{-2}$. We attribute this splitting to a coupling with another molecular energy level induced by the CDT laser.  

The $1\sigma$ statistical error on the fit to each resonance location and to the extrapolated field free resonance location is typically $250~$kHz, which is small compared to the absolute uncertainty of the frequency comb. Since the magnetic field is confirmed to be less than $20~$mG, any systematic shifts due to the residual magnetic field are negligible. Note that at low dipole trap intensities, the CDT potential is significantly tilted due to gravity which leads to a spilling of atoms out of the trap. For our trap geometry, $9.6~$kW~cm$^{-2}$ is the lowest CDT intensity we can use without incurring a large loss of atoms.

\begin{table}
\caption{Experimentally measured PA resonances for $s$-wave collisions in an incoherent mixture of the $\one$ and $\two$ states of $^6$Li.  These PA resonances correspond to a transition from an initial unbound molecular state with $N=0, G=0$ to the $v^{\mathrm{th}}$ vibrational level of the $A^{1}\Sigma_{u}^{+}$ excited state with $N'=1$,$G'=0$. For these measurements, the CDT intensity was 9.6~kW~cm$^{-2}$ and the PA laser intensity was 65~W~cm$^{-2}$.  The absolute uncertainty in each of these measurements is $\pm 600$kHz.  The ac Stark shift of each resonance induced by the PA laser and the CDT laser is also listed, where the number in brackets is an estimation of the 1$\sigma$ errror on the last digit(s). 
}
\begin{ruledtabular}
\begin{tabular}{c c c c}

$v$ & Feature & ODT Shift Rate & PA Shift Rate \\
 & (GHz) &  kHz / (kW/cm$^2$) &  kHz / (kW/cm$^2$)\\
  \hline
29 & 363113.1067 & 199(6) & -745(661) \\
  \hline
30 & 368015.0436 & -546(11) & -2120(783) \\
  \hline
31 & 372780.6714 & 44(4) & -73(228) \\
  \hline
32 & 377406.2393 & -79(6) & 803(707) \\
  \hline
33 & 381887.7859 & 100(5) & 80(253) \\
  \hline
34 & 386221.1190 & 73(5)  & 272(265) \\
  \hline
35 & 390401.8749 & -9(10) & -826(437) \\
\end{tabular}
\end{ruledtabular}
\label{tab:N1-resonances}
\end{table}
%
%%%%%%%%%%%%%%%%%%%%
%% INTERPRETATION %%
%%%%%%%%%%%%%%%%%%%%
%
To interpret our results, we first consider the allowed quantum numbers for the initial and final states of the colliding complex (see \cite{PhysRevA.87.052505} for a more detailed explanation). The initial unbound molecular state can be labelled by $|N,G,f_1,f_2\rangle$ where $\vec{f}_m = \vec{s}_m +\vec{i}_m$ is the total angular momentum of the $m^{\mathrm{th}}$ atom ($m=1,2$). The total spin angular momentum is $ \vec{G} = \vec{f}_1+\vec{f}_2$ and $N$ represents the molecular rotational angular momentum. Particular values of $G$ represent symmetric spin states (specifically, $G=f_1+f_2,f_1+f_2-2,\ldots$) and are associated with even values of $N$, while the remaining possible values of $G$ (that is, $G=f_1+f_2-1,f_1+f_2-3,\ldots$) represent antisymmetric spin states and are associated with odd values of $N$ \cite{PhysRevA.87.052505}. The energy of the colliding complex is given by the sum of the atomic energies, measured with respect to the hyperfine center of gravity (i.e., the energy the complex would have in the absence of hyperfine coupling). For the case of $^6$Li, $E_{\mathrm{F=1/2}} = -a_{2S}^{(6)}$ where $a_{2S}^{(6)} = 152.137~$MHz is the $2S_{1/2}$ atomic hyperfine constant \cite{RevModPhys.49.31}.

The final $A^{1}\Sigma_{u}^{+}$ state can be labelled by $|N,I,G\rangle$ where the total electronic spin ($S=0$) is well defined. Here, $\vec{G} = \vec{I}+\vec{S}$ and $\vec{I} = \vec{i}_1+\vec{i}_2$ is the total nuclear spin of the molecule, which can take on three possible values ($I=0,1,2$) since the nuclear spin of each atom is $i=1$. In the $A^{1}\Sigma_{u}^{+}$ state, odd (even) $N$ levels are (anti)symmetric upon atom exchange \cite{PhysRevA.53.3092}. Thus, similar to the symmetry of $G$, $I=0,2$ corresponds to states with odd $N$ and $I=1$ corresponds to states with even $N$. Table \ref{tab:allowed} shows the possible initial and final states for the system.

In our measurements, $f_1 =f_2 = 1/2$, so for the initial unbound molecule $G=0,1$. However, our measurements are performed on an ultra-cold ensemble and the $p$-wave (and higher order) collisions are greatly suppressed \cite{PhysRevA.87.052505}. This implies that the initial unbound molecular state has $N=0$ (because only $s$-wave collisions occur) and therefore $G=0$ due to symmetry considerations. For dipole transitions between $\Sigma$ states (under consideration here), selection rules state that $\Delta N=\pm1$ and $\Delta G=0$. Thus, there is only one possible transition to the $A^{1}\Sigma_{u}^{+}$ state: $(N=0,G=0) \rightarrow (N'=1,G'=0)$. Finally, the initial energy of the colliding complex is lower than the hyperfine center of gravity by $2a_{2S}^{(6)}$. This extra energy must be added to the $D1$ transition frequency when determing the binding energies of each vibrational level in the $A^{1}\Sigma_{u}^{+}$ state, see Table \ref{tab:binding_energies}. The binding energy thus computed is with respect to the $2^{2}S_{1/2}+2^{2}P_{1/2}$ asymptote.

\begin{table}[h]
\caption{Allowed rotational levels and corresponding nuclear spin configurations for $^{6}\mathrm{Li}_2$ molecules.}
\begin{tabular}{c l l l l}
State &  Electronic & Nuclear & allowed & Total \\
 &  spin & spin & rotational states &  Spin \\
 \hline
 \hline
 \multicolumn{5}{c}{ground states}\\
 \hline

\hline
 - & - & - & $N=0,2,4\ldots$ & $G=0$\\
 - & - & - & $N=1,3,5\ldots$ & $G=1$\\
 \hline

 \hline
 \hline
 \multicolumn{5}{c}{excited states}\\
 \hline

 $A^{1}\Sigma_{u}^{+}$ : & $S=0$ & $I=0$ & $N=1,3,5\ldots$ & $G=0$\\
 & & $I=1$ & $N=0,2,4\ldots$ & $G=1$\\
 & & $I=2$ & $N=1,3,5\ldots$ & $G=2$\\
 \hline

\end{tabular}
\label{tab:allowed}
\end{table}
\begin{table}
\caption{Comparison of the $N'=1$ binding energies found experimentally in this work to the energies predicted by the potential in Table III.A of \cite{LeRoy2009}. In our measurements, the initial unbound molecular state is $2a_{2S}^{(6)}$ below the hyperfine center of gravity of the $2^{2}S_{1/2}+2^{2}S_{1/2}$ threshold. Therefore, the binding energy is computed by adding $2a_{2S}^{(6)}$ to the $D1$ transition energy \cite{PhysRevLett.107.023001} and subtracting the measured photon energy for the PA loss feature. Also listed is the extrapolated field free resonance location, which is used in the determination of the binding energy. All experimentally measured values have an error of $\pm 2\times 10^{-5}~$cm$^{-1}$ ($600~$kHz). All units are cm$^{-1}$. 
}
\begin{ruledtabular}
\begin{tabular}{c c c c c}
    &            &                & This work                       & Ref. \cite{LeRoy2009}\\
$v'$ & Field Free & Binding Energy & (predicted) & (predicted) \\
 & (exp.) &  (exp.) &  Calc.-obs. &  Calc.-obs.\\
  \hline
29 & 12112.14946 & 2791.15743 & 4.07 $\times 10^{-6}$ & -0.0119\\
  \hline
30 & 12275.66069 & 2627.64620 & -1.59 $\times 10^{-5}$ & -0.0121 \\
  \hline
31 & 12434.62472 & 2468.68217 & -1.53 $\times 10^{-5}$ & -0.0123 \\
  \hline
32 & 12588.91710 & 2314.38978 & 1.91 $\times 10^{-5}$  & -0.0126 \\
  \hline
33 & 12738.40534 & 2164.90154 & 2.62 $\times 10^{-5}$ & -0.0129 \\
  \hline
34 & 12882.94978 & 2020.35711  & -7.55 $\times 10^{-6}$ & -0.0132 \\
  \hline
35 & 13022.40482 & 1880.90207 & -1.64 $\times 10^{-5}$ & -0.0134 \\
\end{tabular}
\end{ruledtabular}
\label{tab:binding_energies}
\end{table}
%
%%%%%%%%%%%%%%%%%%%%%%%
%% REFINED POTENTIAL %%
%%%%%%%%%%%%%%%%%%%%%%%
The most accurate potentials for the $X(1^1\Sigma_g^+)$  state of $^{6,6}$Li$_2$, $^{6,7}$Li$_2$ and $^{7,7}$Li$_2$ published to date are those defined according to the reference isotoplogue potential published in \cite{LeRoy2011}, with BOB (Born-Oppenheimer Breakdown) corrections defined according to the $X(1^1\Sigma_g^+)$ BOB correction functions in \cite{LeRoy2009}. The most accurate potentials published for the $A(1^1\Sigma_u^+)$ state of the same isotopologues are those defined according to the reference isotopologue potential and BOB correction functions in \cite{LeRoy2009}. 

The potentials published in \cite{LeRoy2009,LeRoy2011} were optimized to the dataset detailed in Table II of \cite{LeRoy2009}. In this section we discuss a refinement to these potentials, obtained by fitting similar models to a dataset that includes all of the measurements in the dataset described by Table II of \cite{LeRoy2009}, and also the seven new measurements described in this paper. All fits of these models to the data are done with the freely available program \texttt{DPotFit} \cite{d:DPotFit}. 

Since we did not measure new levels for the $X(1^1\Sigma_g^+)$ state, we use precisely the same reference isotoplogue model as in \cite{LeRoy2011} (i.e.~an MLR$_{5,3}^{4.07}(16)$-d model -- the notation for all models is described in \footnote{In this paper we denote the MLR (Morse/Long-range) potential models for the reference isotopologues with the notation MLR$_{p,q}^{r_{\rm{ref}}}(N_\beta)$, or with the notation MLR$_{p,q}^{r_{\rm{ref}}}(N_\beta)$-d if damping funtions are included in $u_{\rm{LR}}$. We denote the model for the adiabatic BOB correction function of atom $A$ by ad-BOB$_{p_{\rm{ad}},q_{\rm{ad}}}(N_u^A)$, and the model for the non-adiabatic BOB correction function of atom $A$ by na-BOB$_{p_{\rm{na}},q_{\rm{na}}}(N_t^A)$), with $u_{\rm{LR}}$ defined according to Eq. 10 of \cite{LeRoy2011}}), and we use precisely the same BOB correction function model as in \cite{LeRoy2009} (i.e.~an ad-BOB$_{6,6}(2)$ model, and no non-adiabatic BOB correction). The parameters of the model for the reference isotoplogue and the models for the BOB correction functions were re-optimized to the new dataset.

For the $A(1^1\Sigma_u^+)$ state, the potential for the reference isotopologue reported in \cite{LeRoy2009} was an MLR$_{6,3}^{4.40}(16)$ with $u_{\rm{LR}}$ defined according to Eq.~20 of \cite{LeRoy2009}. The adiabatic BOB correction function was an ad-BOB$_{3,3}(5)$ model, and the non-adiabatic BOB correction function was a na-BOB$_{3,3}(1)$ model. To refine the potentials for the $A(1^1\Sigma_u^+)$ state, we first used the same BOB correction function models as in \cite{LeRoy2009}, and models for the reference isotopologue of the same form as in \cite{LeRoy2009}, but with $r_{\rm{ref}}$ varying from 3.8 \AA ~ to 5.6 \AA ~ in increments of 0.1 \AA , with $q \in \{2,3\}$ and with $N$ from 9 to 19 \footnote{Although \textit{ab initio} values for higher order dispersion constants are known \cite{Tang2011,Zhang2007}, we only use terms up to $C_8$. This is because in a plot analogous to Fig. 6 of \cite{PhysRevA.87.052505}, we observed that the empirical $A(1^1\Sigma_u^+)$-state potential from Table III.A of \cite{LeRoy2009} starts notably deviating from the theoretical long-range potential (according to \cite{Aubert-Frecon1998}, with the same constants as in Table VI of \cite{PhysRevA.87.052505}, bearing in mind the appropriate symmetry relations) at much larger internuclear distances than the region where the effects of the $C_m, m\ge 9$ terms starts notably deviating from that same theoretical long-range potential. The experimental data on which that empirical potential was based, extended far beyond both of these regions where deviations from the theoretical long-range potential became noticable.}.  After varying $r_{\rm{ref}}$, $q$, and $N$ in this way we found the exact same model as in \cite{LeRoy2009} to be optimal, even with the new data reported in this paper. The MLR$_{6,3}^{4.40}(16)$ had the lowest dimensionless root-mean-square deviation (labeled $\overline{dd}$, and defined by Eq. (2) of \cite{Dattani2011}) out of all of the MLR$_{6,3}^{r_{\rm{ref}}}(16)$ models. MLR$_{6,3}^{r_{\rm{ref}}}(N_\beta >16)$ models never had a $\overline{dd}$ that was more than 0.1\% lower than that of the MLR$_{6,3}^{4.40}(16)$ model, but the MLR$_{6,3}^{4.40}(16)$ model had a $\overline{dd}$ that was lower than the best MLR$_{6,3}^{r_{\rm{ref}}}(15)$ model by more than 1\%. Finally, every MLR$_{6,2}^{r_{\rm{ref}}}(16)$ model had a $\overline{dd}$ that was at least 1\% higher than that of the MLR$_{6,3}^{4.40}(16)$ model, so $N$ would have to be larger than 16 for a $q=2$ model to be competitive with the MLR$_{6,3}^{4.40}(16)$ model. 

After the MLR$_{6,3}^{4.40}(16)$ model was chosen for the reference isotopologue, it was discovered that one of the adiabatic BOB parameters had a 95\% confidence limit uncertainty larger than the parameter itself. This suggests that although the final potentials of \cite{LeRoy2009} had no fitting parameters with 95\% confidence limit uncertainties larger than the parameters themselves, perhaps the ad-BOB$_{3,3}(5)$ model chosen based on \cite{LeRoy2009} has an $N_u^{\rm{Li}}$ value that is larger than necessary for a satisfactory fit. Indeed, reducing $N_u^{\rm{Li}}$ to $N_u^{\rm{Li}}=4$ yielded a fit where none of the $A(1^1\Sigma_u^+)$ state parameters had a 95\% confidence limit uncertainty larger than itself, and $\overline{dd}$ only increased by less than 0.15\%. The na-BOB$_{3,3}(1)$ model that was used in \cite{LeRoy2009} for the non-adiabatic BOB correction function was again found to be optimal with the current dataset.

With the MLR$_{6,3}^{4.40}(16)$ model chosen for the reference isotopologue, and the ad-BOB$_{3,3}(4)$ and na-BOB$_{3,3}(1)$ models chosen for the adiabatic and non-adiabatic BOB correction terms respectively, the SRR (sequential rounding and re-fitting) procedure of \cite{LeRoy1998} was carried out as described in \cite{PhysRevA.87.052505}, yielding a fit with $\overline{dd}$=1.0101. This $\overline{dd}$ value is less than 0.6\% larger than in \cite{LeRoy2011}, despite seven new data points that are 10 times more precise than the most precise data in the dataset of \cite{LeRoy2009}, and despite there being one fewer parameter for the adiabatic BOB correction function. The parameters for the final fit after the SRR procdure are listed in Table \ref{tab:mlr_par}. 

\begin{table}[h!]
\caption{Parameters defining our recommended MLR potentials. Parameters
in square brackets were held fixed in the fit, while numbers in round
brackets are 95\% confidence limit uncertainties from before the SRR procedure, in the last digit(s)
shown. The analysis used the $^{6}$Li$_{2}$ $^{2}P_{\nicefrac{1}{2}}\leftarrow ~ ^2S_{\nicefrac{1}{2}}$
excitation energy of $D1=14903.2967364$ cm$^{-1}$ from \cite{PhysRevLett.107.023001} and the $^6$Li $^{2}P_{\nicefrac{3}{2}}\leftarrow ~ ^2P_{\nicefrac{1}{2}}$
spin-orbit splitting energy of $D2-D1= 0.3353246$ cm$^{-1}$ from \cite{PhysRevA.87.032504}. All $C_m$ values that were held fixed were the non-relativistic \textit{ab initio} values for $^6$Li from \cite{Tang2009}. When numbers taken from the literature were converted into the units used here, they were then rounded to the first digit of their converted uncertainty. Units of length and energy are \AA ~ and cm$^{-1}$ respectively. The polynomial coefficients $\beta_{i}$ for the MLR functions and $t_i$ for the non-adiabatic BOB correction function are dimensionless, while the polynomial coefficients $u_i$ for the adiabatic BOB correction functions have units cm$^{-1}$. For this parameter set, $\overline{dd}=1.0101$.}
\setlength{\tabcolsep}{8pt}
\begin{tabular}{c l c l}
\hline\hline 
& \multicolumn{1}{c}{$X^{1}\Sigma_{g}^{+}$}   &   &\multicolumn{1}{c}{$A^{1}\Sigma_{u}^{+}$}      \\
\hline
$\mathfrak{D}_e$        & 8516.7800\,(23)         &                 &9353.1795\,(28)            \\
$r_e$                   & 2.6729874\,(19)         &                 &3.1079288\,(36)            \\
$C_6$                   & $[6.7190\times 10^6]$   & $ C_3^\Sigma$   &$3.578352\times 10^5$      \\
$C_8$                   & $[1.12635\times 10^8]$  & $C_6^\Sigma$    &$[1.00059\times 10^7]$   \\
$C_{10}$                & $[2.78694\times 10^9]$  &  $C_8^\Sigma$   &$[3.69965\times 10^8]$    \\
$\rho_{\rm{Li}}$                 & [0.54]                    &           & [$\infty$] \\
$\{p,\,q\}$               & \{5,\,3\}                 &                 & \{6,\,3\}                   \\ 
$r_{\mathrm{ref}}$ 				& [4.07] & & [4.4] \\
$\beta_0$     &   0.13904114                  &  &  -1.757571385 \\
$\beta_1$     &  -1.430265                    &  &  -1.0348756   \\
$\beta_2$     &  -1.499723                    &  &  -1.811999    \\
$\beta_3$     &  -0.65696                     &  &  -1.62322     \\
$\beta_4$     &   0.33156                     &  &  -1.44465     \\
$\beta_5$     &   1.02298                     &  &   1.23225     \\
$\beta_6$     &   1.2038                      &  &   2.96072     \\
$\beta_7$     &   1.223                       & &   2.5664      \\
$\beta_8$     &   3.122                       &  &   1.9043      \\
$\beta_9$     &   6.641                       &  &  16.423       \\
$\beta_{10}$  &   0.371                       &  &  20.631       \\
$\beta_{11}$  & -12.17                        &  & -26.932       \\
$\beta_{12}$  &   2.98                        &  & -55.5         \\
$\beta_{13}$  &  28.6                         &  &   8.1         \\
$\beta_{14}$  &   6.8                         &  &  45.6         \\
$\beta_{15} $ & -25.2                         &  &   2.7         \\
$\beta_{16}$  & -15                          &  & -14          \\   
\{$p_{\rm{ad}}$,\,$q_{\rm{ad}}$\}               & \{6,\,6\}    &                 & \{3,\,3\}   \\  
$u_0$         &  0.2210\,(98)                &  &  1.249\,(11)        \\
$u_1$         &  0.18\,(16)                  & &  3.80\,(32)         \\
$u_2$         &  0.54                        &  & -1.09         \\
$u_3$         &  ---                           &  &  3.9          \\
$u_4$         &  ---                             &  & -6.4          \\
$u_\infty$    & [0]                          &  & [1.2315155]   \\
\{$p_{\rm{na}}$,\,$q_{\rm{na}}$\}               & ---    &                 & \{3,\,3\}   \\ 
$t_0$         &  ---                 & &  [0]         \\
$t_1$         &  ---                        &  & 0.000109         \\
$t_\infty$         &  ---                           &  &  [0]          \\ 
\hline
\end{tabular}

\label{tab:mlr_par}
\end{table}

In conclusion, we have performed high resolution photoassociation spectroscopy of the $^{6}$Li$_2$ $A(1^1\Sigma_u^+)$ state in a degenerate Fermi gas of  $^{6}$Li atoms. We have observed seven vibrational levels $v=29-35$ and determined the position of each level to an absolute uncertainty of $\pm 0.00002$~cm$^{-1}$ ($\pm 600$ kHz). We use these data to further refine the analytic potential energy function for this state and provide the updated Morse/Long-Range model parameters. 

These measurements take place in the broader context of studying and characterizing the ground and excited state potentials of $^{6}$Li$_2$ molecules. Specifically, they serve as ideal intermediate levels for: (a) a high resolution study of the ground-singlet $X(1^1\Sigma_g^+)$ state in the ultra-cold regime, (b) a refinement and improvement on the uncertainty of the $s$-wave scattering length in  $^{6}$Li and (c) as a starting point for the creation of ground state molecules using a two-photon STIRAP process.

We gratefully acknowledge Takamasa Momose for the use of the Ti:sapphire laser.  We also thank Robert J Le$\,$Roy for many helpful discussions.  The authors also acknowledge financial support from the Canadian Institute for Advanced Research (CIfAR), the Natural Sciences and Engineering Research Council of Canada (NSERC / CRSNG), and the Canadian Foundation for Innovation (CFI).  N.S.D. also thanks the Clarendon Fund for financial support.
%\begin{thebibliography}{99}
%\bibliographystyle{plain}
\bibliography{Li_PA}

\end{document}